\documentclass[runningheads]{llncs}
\usepackage[T1]{fontenc}
\usepackage{subcaption}

\usepackage{graphicx}
\usepackage{hyperref}
\usepackage{amsmath}
\usepackage{algorithm}
\usepackage{algorithmic}
\usepackage{geometry}
\usepackage{setspace}

\usepackage[table,xcdraw]{xcolor}
\usepackage{booktabs}
\usepackage{multicol}

\usepackage{xcolor}

\usepackage{bbding}

\begin{document}
\title{MeSH Concept Relevance and Knowledge Evolution: A Data-driven Perspective}
%
%
\author{Jenny Copara\inst{1}\Envelope\orcidID{0000-0002-1510-3331} \and
Nona Naderi\inst{2}\orcidID{0000-0002-1272-7640} \and
Gilles Falquet\inst{1}\orcidID{0000-0002-1401-5110}
\and
Douglas Teodoro\inst{1}\orcidID{0000-0001-6238-4503}}
\authorrunning{J. Copara et al.}
%
\institute{University of Geneva, Geneva, Switzerland \\
\email{jenny.copara@etu.unige.ch, gilles.falquet@unige.ch, douglas.teodoro@unige.ch}\\
\and
Université Paris-Saclay, CNRS, Laboratoire Interdisciplinaire des Sciences du Numérique, París, France \\
\email{nona.naderi@universite-paris-saclay.fr}\\
}
\maketitle              
\begin{abstract}
The Medical Subject Headings (MeSH), one of the main knowledge organization systems in the biomedical domain, continuously evolves to reflect the latest scientific discoveries in health and life sciences. Previous research has focused on quantifying information in MeSH primarily through its hierarchical structure. In this work, we propose a data-driven approach based on information theory and network analysis to quantify the relevance of MeSH concepts. Our method leverages article annotations and their citation networks to compute four aspects of relevance --- informativeness, usefulness, disruptiveness, and influence --- over time. Using both the citation network and the MeSH hierarchy, we compute these relevance aspects and apply an aggregation algorithm to propagate scores to parent nodes. 
We evaluated our approach on MeSH terminology changes and showed that it effectively captures the evolution of concepts. 
The mean relevance of evolving concepts is higher compared to concepts that remained unchanged ($2.09E-03$ \textit{vs.} $8.46E-04$). Moreover, we validated the framework by analyzing retracted articles and found that concepts used to annotate retracted articles (mean relevance: 0.17) differ substantially from those annotating non-retracted ones (mean relevance: 0.15). Overall, the proposed framework provides an effective method for ranking concept relevance and can support the maintenance of evolving knowledge organization systems.

\keywords{terminology evolution \and concept relevance \and information theory \and network analysis}
\end{abstract}
%
%
%
\section{Introduction}
The vast amount and complexity of the scientific biomedical literature require sophisticated methods of structuring and reasoning of knowledge~\cite{Ivanovic2014,Wang2023}. Knowledge organization systems (KOSs) --- including terminologies, thesauri, and ontologies --- are designed to formally encode concepts,  enhancing discoverability, representation, and reasoning across biomedical corpora~\cite{Ivanovic2014}. Among these KOSs, the Medical Subject Headings (MeSH) plays a central role by indexing, structuring, and facilitating search within the leading biomedical corpus, MEDLINE, which is accessible through the PubMed interface~\cite{Lipscomb2000}. 

MeSH is a controlled vocabulary maintained by the U.S. National Library of Medicine~\cite{Lowe1994}. Functionally, it is a thesaurus encompassing several types of concepts~\cite{Nentidis2021,Ivanovic2014,nlm}. The primary units for indexing and retrieval are descriptors, which are used to annotate the MEDLINE/PubMed dataset. Additional elements, such as qualifiers and supplementary concept records, provide further specification and context for descriptors. MeSH headings represent biomedical concepts and serve as standardized interpretations for groups of related medical terms~\cite{Ivanovic2014,nlm}. MeSH headings are organized hierarchically into sixteen main categories, and descriptors map directly to their appropriate positions in the hierarchy~\cite{Ivanovic2014}. Beyond facilitating search and question answering in the biomedical literature~\cite{gobeill2009question,Gobeill2009TakingBO}, MeSH also supports the study of structural and evolutionary patterns in biomedicine. Furthermore, this thesaurus assists in the evaluation of research grants and helps identify emerging research topics of high interest~\cite{Fortunato2018,Ilgisonis2022,Wu2019}.

As knowledge in the biomedical domain evolves, MeSH needs to be periodically modified and updated to reflect these changes~\cite{Lu2022,Nelson2007}. Based on the current knowledge state, MeSH concepts (i.e., headings) encode a certain amount of information, which may vary as knowledge develops~\cite{Balogh2019}. The relevance of some concepts may decrease or increase, rendering them obsolete, necessitating drastic changes or additions~\cite {Nentidis2021,Balogh2019,Cardoso2018}. As a consequence, the hierarchy structure can change. Thus, it is crucial to identify the most representative MeSH concepts~\cite{Ivanovic2014,Konopka2015} to summarize the content of the thesaurus, prioritize maintenance tasks in the KOS, and improve the performance of information retrieval systems that rely on this KOS, such as PubMed~\cite{Cardoso2020,Cardoso2018,Fernandez2021}. 

Researchers have applied different approaches to quantify the information encoded in MeSH by analyzing features such as the hierarchy, bibliometric indices of articles, and the co-occurrence of descriptors~\cite{Nentidis2021,Balogh2019,Chien2019,Ilgisonis2022,Kastrin2019}.  
Balogh \textit{et al.}~\cite{Balogh2019} showed that nodes in this thesaurus with many children tend to acquire more children over time. 
Chien \textit{et al.}~\cite{Chien2019} examined PubMed articles in medicine and health using the x-index. Ilgisonis \textit{et al.}~\cite{Ilgisonis2022} analyzed descriptors to investigate trends in general and personalized medicine. Several studies examined descriptor co-occurrence, particularly in relation to new descriptors~\cite{Nentidis2021,Kastrin2019}. Kastrin and Hristovski~\cite{Kastrin2019} built a network from descriptor co-occurrences to investigate indexing trends. Anastasios \textit{et al.}~\cite{Nentidis2021} demonstrated that approximately one-quarter of new descriptors in MeSH represent emerging concepts. 
Ilgisonis \textit{et al.}~\cite{Ilgisonis2022} further focused on related descriptors, defined as descriptors that co-occur with annotated descriptors. Despite these contributions, the analysis of instances—such as the occurrence of descriptors in PubMed articles—remains largely unexplored for quantifying MeSH concepts over time.

This study addresses this gap by proposing an approach to measure the relevance of MeSH concepts through their instantiation in biomedical literature. The approach integrates three main factors: the usage of concepts in PubMed articles, the hierarchical structure of MeSH, and the citation network of PubMed articles, analyzed through information-theoretic and network-based methods. We define four dimensions of relevance: informativeness, usefulness, disruptiveness, and influence. To quantify informativeness, we apply entropy from information theory~\cite{Ghahramani2006,Shannon1948}. To measure usefulness, we use category utility supported by the hierarchical organization of MeSH~\cite{Corter1992}. To assess disruptiveness, we consider whether concepts annotate highly cited but consolidated technologies, which may not contribute novelty to the field~\cite{Figueiredo2019,Funk2017,Wu2019}. To evaluate influence, we apply centrality metrics from graph theory~\cite{Newman2018}, which capture the importance of nodes in networks~\cite{Hamilton2020}. We apply these methods to the MEDLINE database and analyze MeSH annotations over time. We generate monthly data by mapping references to MeSH concepts and computing each metric for each concept. We propagate this data through the MeSH hierarchy and combine the multidimensional scores using reciprocal rank fusion (RRF) to estimate the relative relevance of MeSH concepts.

This paper presents three main contributions: 
\begin{enumerate}
\item We propose an algorithm to compute the relevance of MeSH concepts by propagating information from a network of instances to the KOS structure. The network of MeSH instances is based on the PubMed citation network, while the knowledge structure relies on the thesaurus hierarchy. 
\item We assess the relevance of MeSH concepts quantified by multidimensional metrics --- informativeness, usefulness, disruptiveness, and influence --- and show how they vary over time for different hierarchical levels.
\item We provide a quantitative evaluation of the approach in two use cases: MeSH terminology evolution and analysis of retracted papers. We compare the relevance of MeSH terms in a test group (concept changes and retracted papers) versus a control group (no evolution and no retraction, respectively).
\end{enumerate}

\section{Materials and methods}
We explore the evolution of MeSH by analyzing concept instantiations across multiple years in PubMed. For this purpose, we use the PubMed/MEDLINE corpus, which contains articles manually annotated with MeSH descriptors. In this section, we present the methodology used to create our dataset. We also describe the metrics used to quantify relevance in a KOS. Next, we discuss strategies to propagate these metrics through the hierarchy and to compute a metrics fusion. Finally, we outline the evaluation criteria applied in this work.

\subsection{Dataset and pre-processing}
\label{subsection:dataset}

We conducted our study using the 2022 collection of PubMed\footnote{\url{https://lhncbc.nlm.nih.gov/ii/information/MBR/Baselines/2022.html}}, which comprises 33,403,054 articles published up to December 2021. PubMed indexes new biomedical and life science articles as soon as they are published. Articles in PubMed are annotated by domain experts using MeSH descriptors, which are organized hierarchically into 16 categories denoted by single letters.

For this collection, we built citation networks using PubMed article citations and the iCite\footnote{\url{https://nih.figshare.com/articles/dataset/iCite_Database_Snapshot_2022-04/19763386}} dataset~\cite{iCite2022} as a more comprehensive source of citation data. We then created cumulative monthly networks, i.e., incorporating nodes from previous months, from January 2014 through December 2021. To build the corresponding MeSH instance hierarchy, we used all the MeSH descriptors assigned to articles in a particular month and mapped them to their respective MeSH tree codes. The first monthly citation network (January 2014) contained 18,728,651 nodes and 381,296,042 edges, while the last one (December 2021) contained 27,728,971 nodes and 673,415,629 edges. Because computing certain metrics was prohibitively time-consuming, we randomly sampled 10\% of the nodes from each original monthly citation network. Edges were retained only when both nodes appeared in the sampled network. Thus, with this approach, we preserve the structure similar to the full citation network. This resulted in the first sampled citation network (January 2014) having 1,872,866 nodes and 3,807,856 edges, while the last sampled network (December 2021) had 2,772,898 nodes and 6,745,896 edges. 

\subsection{MeSH relevance quantification}

\begin {figure}[!h]
\centering
\includegraphics[width=1\textwidth]{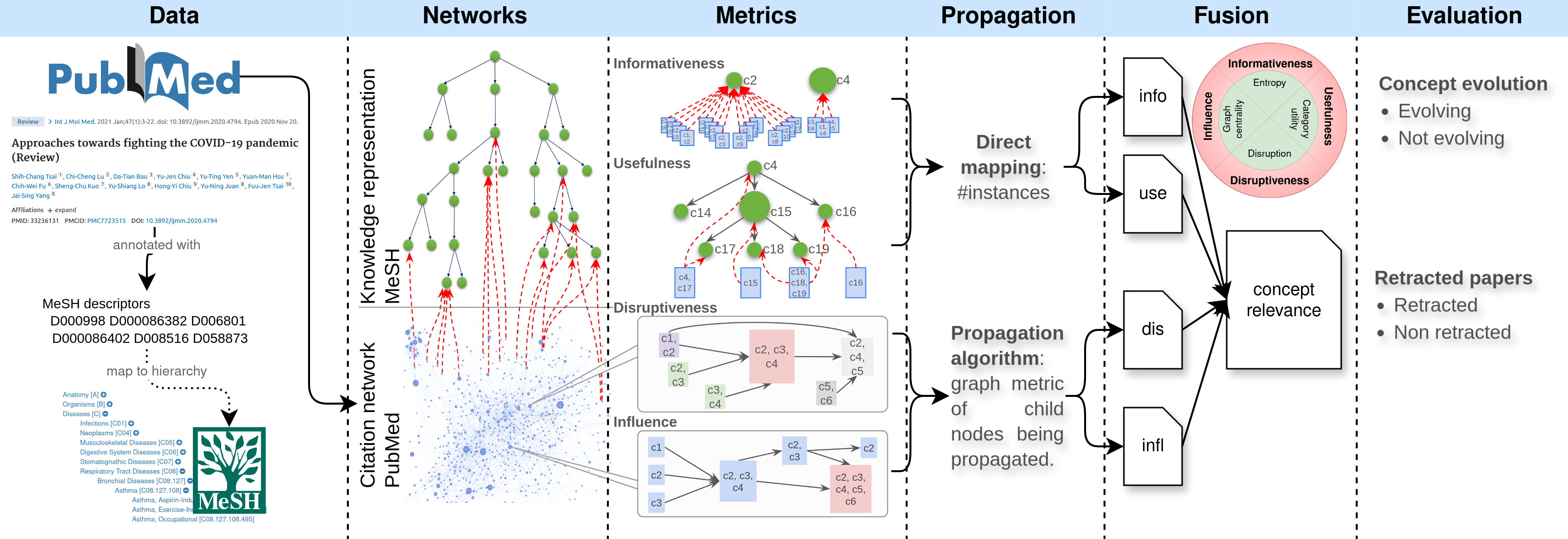}
\caption{
Pipeline of concept relevance quantification in MeSH.
} 
\label{figure:aspects_metrics}
\end{figure}

Analogous to the definition of \textit{relevance} in information science and information retrieval~\cite{Mizzaro1997}, we define the relevance of a concept as the extent to which a concept or a set of concepts meets the representational needs of a dataset~\cite{Schamber1990}. Following the definition of relevance provided by Saracevic~\cite{Saracevic1975}, in our study, we consider multiple aspects to quantify the relevance of MeSH concepts, namely, informativeness, usefulness, disruptiveness, and influence (Figure \ref{figure:aspects_metrics}, Metrics). We provide the code to compute these aspects\footnote{\url{https://github.com/jcoparaz/computation_metrics}} as well as the computed metrics\footnote{\url{https://zenodo.org/records/13635979}}. Each PubMed article is annotated with MeSH descriptors, which map to one or more tree nodes. In the following, we describe these metrics and explain how the relevance of a MeSH concept is quantified according to them.

\subsubsection{Informativeness}
To quantify the relevance of a concept in terms of informativeness, we use entropy, introduced by Shannon~\cite{Ghahramani2006,Shannon1948}. Entropy quantifies the level of uncertainty of a variable. In our case, if a concept is frequently used to annotate many papers, it is less informative, since numerous other articles also contain this annotation. For example, the concept ``Humans'' (D006801) is used to annotate around 70\% of PubMed articles. Therefore, this concept contributes little information when it appears in an article and has a low entropy value. In contrast, the concept ``Proteins'' (D011506) is used to annotate only 2\% of PubMed articles, which makes it considerably more informative.

As shown in Equation \ref{eq:entropy}, we follow Shannon’s entropy definition $H(X)$, where $p(x)$ represents the probability of an event's occurrence and $x$ refers to all values that a variable $X$ can take: 

\begin{equation}
    H(X) = - \sum_x p(x) \log_2 p(x).
    \label{eq:entropy}
\end{equation}
In our study, the instantiation of a MeSH concept represents an event, that is, articles mapped to nodes in the MeSH hierarchy, representing the concept's usage (dotted red arrows in Figure \ref{figure:aspects_metrics}, Networks). The probability of an event $p(x)$ is calculated as the probability that a concept $x$ is mapped within all mappings in its hierarchical level (its distance to the root node). Mappings at lower levels are propagated to their parents when calculating the latter's probability.

\subsubsection{Usefulness}
To quantify relevance in terms of usefulness, we use the category utility measure to identify basic-level categories in a taxonomy~\cite{Corter1992}. Basic-level categories are preferred in both natural and artificial hierarchies, and they carry the most information about the category’s attributes~\cite{Corter1992,Rosch1976} (e.g., when shown an image of a specific oncological exam, most professionals identify it as cancer rather than as a general disease or a colon cancer). A node in a hierarchy is useful to the extent that it improves the ability to predict the features of instances in that category. As shown in Figure \ref{figure:aspects_metrics}, in this study, the mapping of articles to the hierarchy characterizes concepts. Therefore, a concept within a densely mapped branch in the hierarchy is considered more useful (e.g., as node \textit{c15} in the example).

Category utility combines the frequency of a category's occurrence with the probability of its features~\cite{Corter1992}. Let us define a category $c$ as a finite set of instances: \(c = \{o_1, o_2, ..., o_n\}\), where each instance $o_i$ can be described by a finite set of discrete features \(F = \{f_1, f_2, ..., f_m\}\). The category utility (CU) of category $c$ with a set of features $F$ can then be calculated as in Equation \ref{eq:cu}, where $p(c)$ is the probability of a category (i.e., a MeSH concept in our case), $p(f_k)$ is the probability of feature $f_k$ and $p(f_k|c)$ is the conditional probability of feature $f_k$ given category $c$:

\begin{equation}
    CU(c,F) = p(c) \sum_{k=1}^{m} [p(f_k|c)^2 - p(f_k)^2].
    \label{eq:cu}
\end{equation}
To calculate category utility for the nodes in the MeSH hierarchy, we use mapping to articles $a_i$ as the set of features \(F = \{a_1, a_2, ..., a_m\}\), where each feature $a_i$ has a binary value (annotated or not with a category). To compute the probabilities, we created a binary matrix $M_{ij}$ of $n$ tree nodes and $m$ features as follows:

\begin{equation*}
    M_{ij}=
    \begin{cases}
      1, & \text{tree node $i$ is mapped by article $j$},\\
      0, & \text{otherwise}.
    \end{cases}
\end{equation*}

\subsubsection{Disruptiveness}
To measure the relevance of a concept in terms of the aspect of disruptiveness, we use the disruption index~\cite{Figueiredo2019}. This index was originally designed to quantify the extent to which an invention (a concept, in our case) consolidates or destabilizes the subsequent use of the components (i.e., referenced concepts) on which it builds~\cite{Figueiredo2019}. To compute the disruption index $D$ for a focal node, we consider the citation network as a tripartite graph $G = (V_1, V_2, V_3, E)$, where the set of nodes ${V_1, V_2, V_3}$ represents the articles, which are connected by directed edges $E$, representing their citations~\cite{Hamilton2020,Newman2018}. Following Figueiredo and Andrade's definition, given a focal node $a$ (light red in Figure \ref{figure:aspects_metrics}, Metrics, Disruptiveness), corresponding to an article for which we want to compute the level of disruptiveness, there are $j$ articles that reference  $a$ and at least one of predecessor (purple box), $i$ articles that reference $a$ but none of its predecessors (green box), and $k$ articles that do not reference $a$ but reference at least one predecessor (dark gray box), disruption $D$ for $a$ is defined by Equation \ref{eq:disruption}. D can take values in the range [-1,1].

\begin{equation}
    D=\frac{n_i-n_j}{n_i+n_j+n_k},
    \label{eq:disruption}
\end{equation}
\noindent
where $n_i$, $n_j$, and $n_k$ represent the number of $i$, $j$, and $k$ articles, respectively. The disruption score $D$ of an article is then propagated to the concepts mapped to it ($c2$, $c3$, and $c4$ in the example). Finally, the disruption index $D_c$ for a concept $c$ is the sum over the disruption scores for all articles $a_1$, $…$, $a_n$ which are mapped to $c$, divided by the number of articles in the citation network, i.e., $D_c = \sum_{i=1}^n D_i/m$, where $D_i$ is the disruption index of the $i$\textit{th} article mapped to $c$ and $m$ is the number of articles in the citation network. To compute $D$, we used Figueiredo and Andrade's disruption implementation~\cite{Figueiredo2019}. 

\subsubsection{Influence}
In graph theory, the influence of a node is quantified through graph centrality, based on the importance of neighboring nodes. We use this concept to measure the influence aspect of a MeSH concept. Rather than relying on the MeSH hierarchy, which captures only ontological relations, we use instances (i.e., mapped articles) to identify the influence of MeSH concepts. Similar to the disruptiveness aspect, we consider the citation network to compute centrality.  In Figure \ref{figure:aspects_metrics}, we show a small citation network example to demonstrate how nodes (articles) influence their neighbors. An article annotated with one or more concepts, denoted by $c_1$, $...$, $c_n$, gains influence from its incoming articles (e.g., light red box).

In our work, we used the PageRank algorithm to compute the centrality of a node~\cite{Brin1998,Newman2018}, which uses the importance of the node’s neighbors to compute the node's importance. The PageRank of node $i$ in a directed graph $G$ is defined as $x_i$ in Equation \ref{eq:pr}:
\begin{equation}
    x_i=\alpha \sum_j A_{ij} \frac{x_j}{k_j^{out}}+\beta ,
    \label{eq:pr}
\end{equation}
\noindent
where the parameter $\alpha$ regulates the contribution between the eigenvector and the constant term $\beta$. The term $\beta$ is used to initialize nodes with zero out-degree. The adjacency matrix of $G$ is denoted by $A_{ij}$, which captures the edges between pairs of nodes (citations in our case). The out-degree of node $j$ is represented by $k_j^{out}$. Similar to disruptiveness, the centrality $x_i$ of an article $i$ is propagated to the concepts it is mapped to, and the overall centrality $x_c$ of a concept is computed by summing over all the centralities of articles that concept $c$ is mapped to divided by the number of articles in the citation network, i.e., $x_c = \sum_{i=1}^n x_i/m$, where $x_i$ is the centrality score of the $i$\textit{th} article mapped to $c$ and $m$ is the number of articles in the network. We used the PageRank implementation from the NetworKit library\footnote{\url{https://networkit.github.io/}} to calculate the graph centrality, where $\alpha = 0.85$ and $\beta=1-\alpha$.

\subsection{Propagation of data instances to the knowledge representation}

We encoded the usage of MeSH instances and propagated them to the knowledge representation for each aspect. Given the “\textit{broader than}” relationship
between MeSH concepts, we consider two methods for propagating information to higher-level nodes: direct mapping or a cumulative weighted propagation algorithm. For the information-theoretical measures (entropy and category utility), which are computationally efficient, we directly propagate instance mappings to their parent concepts using the tree nodes associated with each descriptor. Then, we compute the relevance scores for each node in the hierarchy. 

We approach the network analysis metrics (disruption and graph centrality) differently due to their high computational complexity. Instead of combining the mappings into higher-level nodes, we aggregate the scores of lower-level nodes into those of their parents. The assumption is that if a node is disruptive or influential, so too will its parent. 
As described in Algorithm \ref{algo:propagation}, the propagation method requires the MeSH hierarchy $H$ and the network-based relevance scores computed using the citation network (variable $tnm$) as input. In line 2, we obtain tree nodes per level of specificity using the hierarchy $H$; for instance, tree nodes C and D are in level 1, while tree nodes C01 and D01 are in level 2. Then, we iterate over the hierarchy from bottom to top, using a reversed level list (variable $llr$, line 3) to propagate the relevance scores of the lower-level concepts to the higher levels. First, at each level, we get a set of unique parents (lines 5-9). Then, we retrieve their children (line 11) and, utilizing their initial measurements (lines 14-18), compute the weighted parent relevance, variable $gmp$, based on the children's relevance and the number of nodes in their level  (lines 19 and 21). Finally, the parent value is calculated by taking into account the direct mapping to its node in addition to the disruption of child nodes being propagated $d_{n_i} = direct\_d_{n_i}+indirect\_d_{n_i}$ (lines 22 and 23).

\begin{algorithm}[hbt]
\caption{Propagation of graph-based metrics to the MeSH hierarchy}
\begin{multicols}{2}
\begin{algorithmic}[1]
\setstretch{1.2}
\REQUIRE $H$ (hierarchy), $tnm$ (graph metric values initialized from descriptors)
\ENSURE $gmh$ (graph metric of tree nodes in the hierarchy)
\STATE $gmh \gets \{\}$
\STATE $dpt \gets get\_nodes\_in\_levels\_specificity(H)$
\STATE $llr \gets \text{sorted list of keys in } dpt \text{, reversed}$
\FOR{each $current\_level$ in $llr$}
    \STATE $unique\_parents \gets \text{an empty set}$ 
    \FOR{each $treenode$ in $dpt[current\_level]$}
        \STATE $parent \gets \text{get\_parent}(H, treenode)$
        \STATE add $parent$ to $unique\_parents$
    \ENDFOR
    \FOR{each $parent$ in $unique\_parents$}
        \STATE $children \gets \text{get\_children($H$, $parent$)}$
        \STATE $gmp \gets 0$
        \FOR{each $child$ in $children$}
            \IF{$child \in tnm$}
                \IF{$child$ has no children}
                    \STATE $gmh[child] \gets tnm[child]$
                \ENDIF
            \ENDIF
            \STATE $gmp \gets gmp + gmh[child]$
        \ENDFOR
        \STATE $gmp \gets gmp / \text len(dpt[current\_level]$)
        \STATE $dgm \gets tnm[parent] \text{ if } parent \in tnm \text{ else } 0$
        \STATE $gmh[parent] \gets gmp + dgm$
    \ENDFOR
\ENDFOR
\RETURN $gmh$
\end{algorithmic}
\end{multicols}
\label{algo:propagation}
\end{algorithm}

\subsection{Consolidating the relevance ranking of MeSH codes}

Each of the metrics described previously measures different relevance aspects of a MeSH concept according to its data instances. To have a unified relevance score, we propose fusing the multidimensional aspects into a single ranking. Inspired by ranking fusion methods in information retrieval, we consider that each concept $d \in D$, where $D$ is the set of MeSH codes, is ranked according to a relevance dimension $r \in R$, where $R$ is the set of four aspects considered in this work: informativeness, usefulness, disruptiveness, and influence.

Information retrieval research proposes various techniques for merging rankings, such as CombSUM, CombMNZ, CombANZ, rank-biased precision, and reciprocal rank fusion (RRF)~\cite{Cormack2009}. Due to its simplicity, effectiveness, and robustness ~\cite{Cormack2009,teodoro2021information}, we use RRF to compute the overall relevance of a MeSH concept based on its instantiations. RRF combines the outcomes of different information retrieval methods with varying relevance scores into a single result set using the rank information. The mathematical definition of RRF is provided in Equation \ref{eq:rrf}, where, in our case, a concept $d$ belongs to the set of MeSH codes $D$, and $r(d)$ represents the rank of a code $d$ according to a metric $r$ from the set of metrics $R$. The constant $k$ minimizes the impact of outliers. We set $k$ to $60$, as defined in the original RRF paper\cite{Cormack2009,teodoro2021information}.

\begin{equation}
    \text{RRF}(d \in D)=\sum_{r \in R} \frac{1}{k+r(d)} ,
    \label{eq:rrf}
\end{equation}

\subsection{Evaluation criteria}
We qualitatively analyze the trends of MeSH concepts according to the different relevance aspects for various levels of the hierarchy and across MeSH versions from 2014 to 2021. To quantitatively evaluate the effectiveness of our approach, we conducted two experiments. First, we assess the relevance of concepts that changed against those that remained unchanged between MeSH versions. Second, we compare the relevance of concepts between retracted and non-retracted papers in PubMed. We used the Mann-Whitney test to compare relevance changes; \textit{p-values} below $0.05$ are considered statistically significant.


\section{Results}

This section presents results assessing the relevance of MeSH concepts using a multi-aspect model. It includes an analysis of the evolution of the MeSH terminology, highlighting key concepts at different hierarchy levels. Two use cases are explored: i) comparing the relevance of concepts that changed versus those that remained unchanged, and ii) comparing concepts used to annotate retracted papers versus those used for non-retracted papers.

\subsection{MeSH concept relevance trends}

\begin{figure*}[hbt]
\centering
\includegraphics[width=0.80\linewidth]{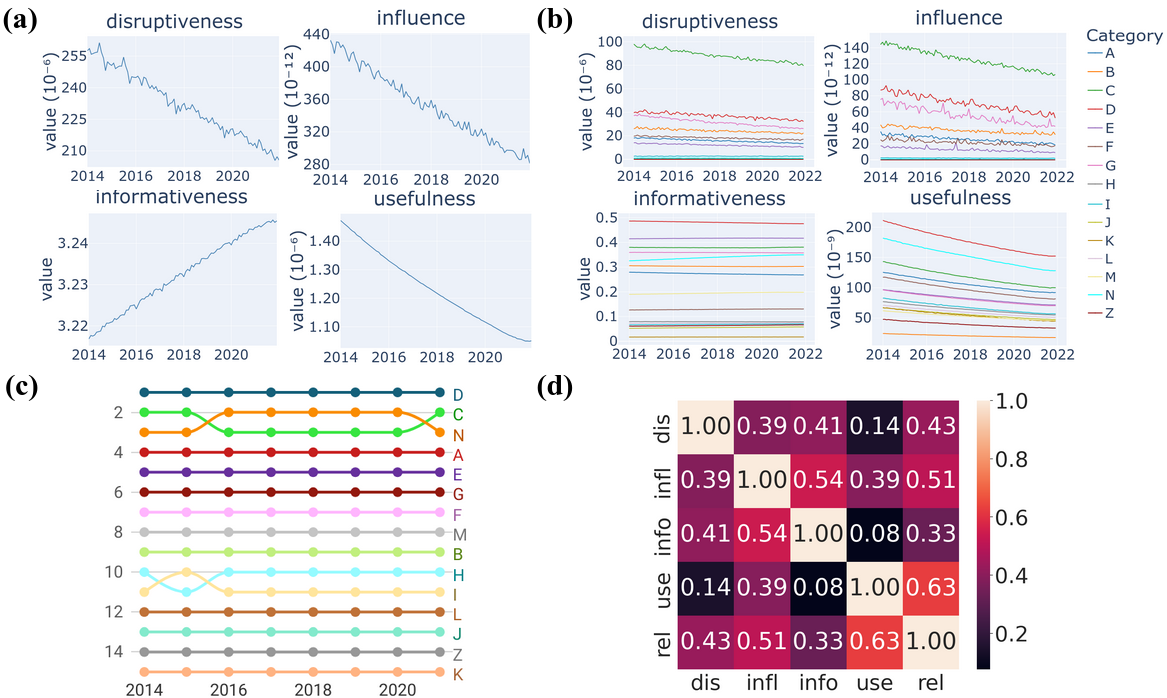} 
\caption{Trend of concept relevance in MeSH at level 1 and correlation of descriptors. a) The sum of MeSH categories. b) Trend of MeSH categories. c) Ranking of concepts at level 1. d) Correlation of aspects. Column names are dis: disruptiveness (disruption), infl: influence (centrality), info: informativeness (entropy), use: usefulness (category utility), rel: concept relevance (RRF fusion). 
} 
\label{figure:panel_metrics_trend_rank_correlation_CDE}
\end{figure*}

We calculated the relevance of MeSH concepts from 2014 to 2021 using the PubMed 2022 snapshot annotated with the MeSH 2022 version. Relevance scores for the different aspects --- disruptiveness, influence, informativeness, and usefulness --- as well as for the overall concept relevance were computed monthly. Figure \ref{figure:panel_metrics_trend_rank_correlation_CDE}a depicts the cumulative trends, i.e., the sum of the first level of the MeSH hierarchy (A, ..., Z).
Disruptiveness, influence, and usefulness decrease over time ($p < 0.001$). Park and colleagues also corroborate the decreasing trend of disruptiveness, although their analysis was based on scientific articles, in contrast to ours, which is based on MeSH concepts~\cite{Park2023}. On the other hand, informativeness, which measures the degree of uncertainty, tends to increase over time ($p < 0.001$). Therefore, the degree of uncertainty also increases as more recent knowledge is added to PubMed.

Figure \ref{figure:panel_metrics_trend_rank_correlation_CDE}b shows the trend of the concept relevance at the first level in MeSH. 
All of the categories show a decreasing trend ($p < 0.001$). The most useful concepts (Figure \ref{figure:panel_metrics_trend_rank_correlation_CDE}b - bottom right) in the first level are D (Chemicals and Drugs), followed by N (Health Care) and C (Diseases), indicating they encode the basic-level categories of PubMed. 
In influence, all categories show a decreasing trend ($p < 0.001$). Concept C is the most influential (Figure \ref{figure:panel_metrics_trend_rank_correlation_CDE}b - top right), indicating that `Diseases' is a central concept in PubMed. It is followed by the concepts D and G (Phenomena and Processes), which include chemical, and physiological phenomena descriptors.

Trends in disruptiveness are mostly decreasing ($p < 0.001$), though categories M and H are not statistically significant. J shows an increasing trend ($p < 0.001$). On the other hand, K shows a growing trend, while I shows no significant trend.
Concept C is also the most disruptive in the first level of the MeSH hierarchy (Figure \ref{figure:panel_metrics_trend_rank_correlation_CDE}b - top left), being more than twice as disruptive as the second-ranked concept (D). 
Trends of informativeness are primarily increasing with a $p < 0.001$. However, categories H, G, B, A, and D present a decreasing trend ($p < 0.001$). Category C shows no trend (not statistically significant). Concept D has the highest informativeness score (Figure \ref{figure:panel_metrics_trend_rank_correlation_CDE}b - bottom left), followed by concept E (Analytical, Diagnostic and Therapeutic Techniques, and Equipment). However, its score gradually decreases, with a standard deviation of 0.003377. The trends of this aspect remain relatively constant over time, largely due to the narrow range of values across  MeSH categories.

Figure \ref{figure:panel_metrics_trend_rank_correlation_CDE}c shows the ranking of concepts in level 1. Most concepts retain their positions over time, except for Diseases (C), Health Care (N), Disciplines and Occupations (H), and I, which show some variation. Chemicals and Drugs (D) consistently ranks higher than other MeSH categories. The second position in this ranking alternates between `Diseases' and `Health Care', with the latter prevailing in five of eight years.

Figure \ref{figure:panel_metrics_trend_rank_correlation_CDE}d shows the correlation between relevance aspects for each descriptor using the monthly values in releases. The overall concept relevance, as a fusion of disruptiveness, influence, informativeness, and usefulness, shows a moderate correlation. Usefulness has a low correlation with disruptiveness and informativeness and a moderate correlation with influence. Influence and informativeness exhibit a moderate correlation.

\subsection{Top concepts in the MeSH terminology}

\begin{figure}[!h]
\centering
\includegraphics[width=0.80\linewidth]{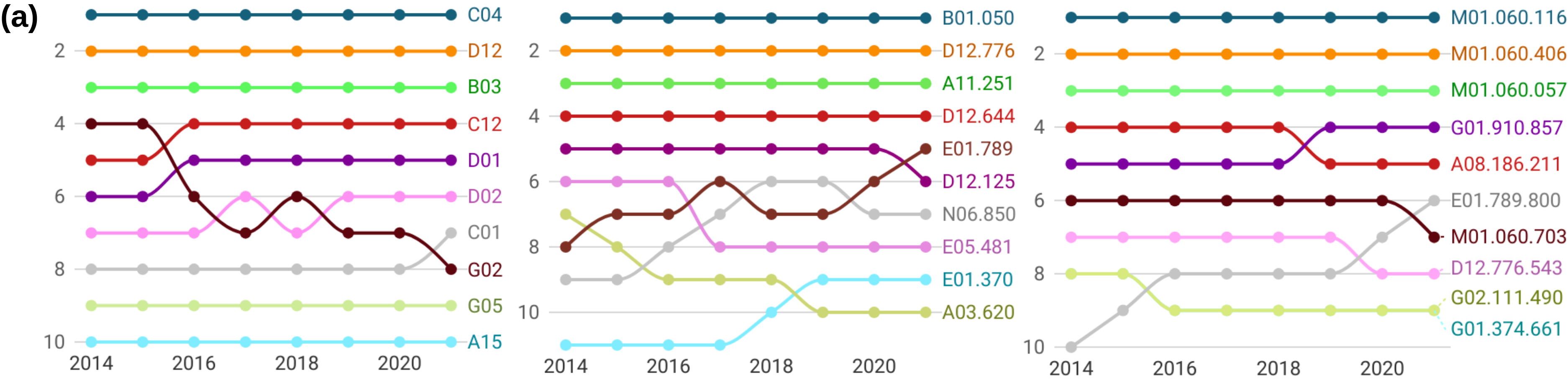} 
\includegraphics[width=0.80\linewidth]{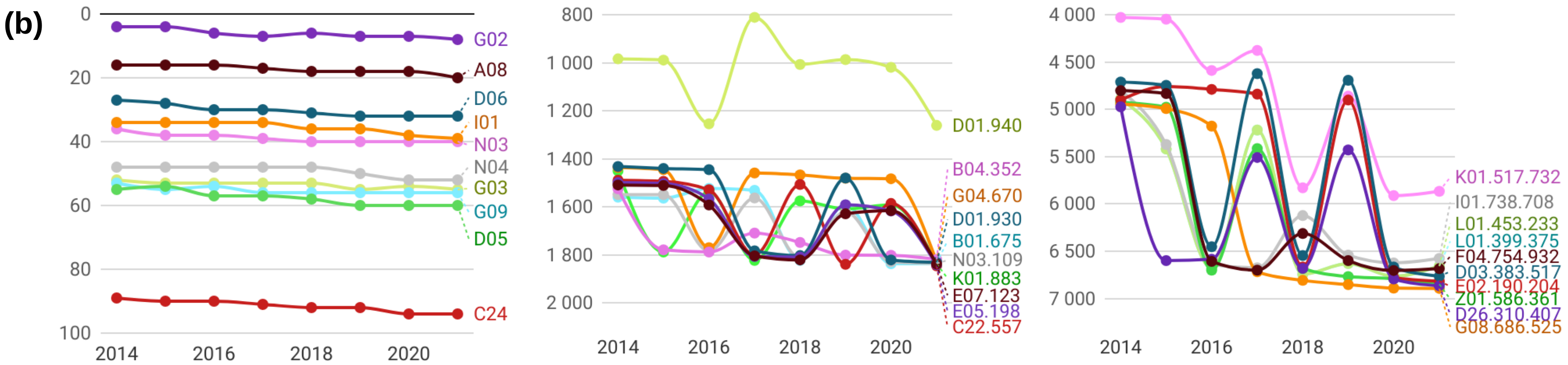}
\includegraphics[width=0.80\linewidth]{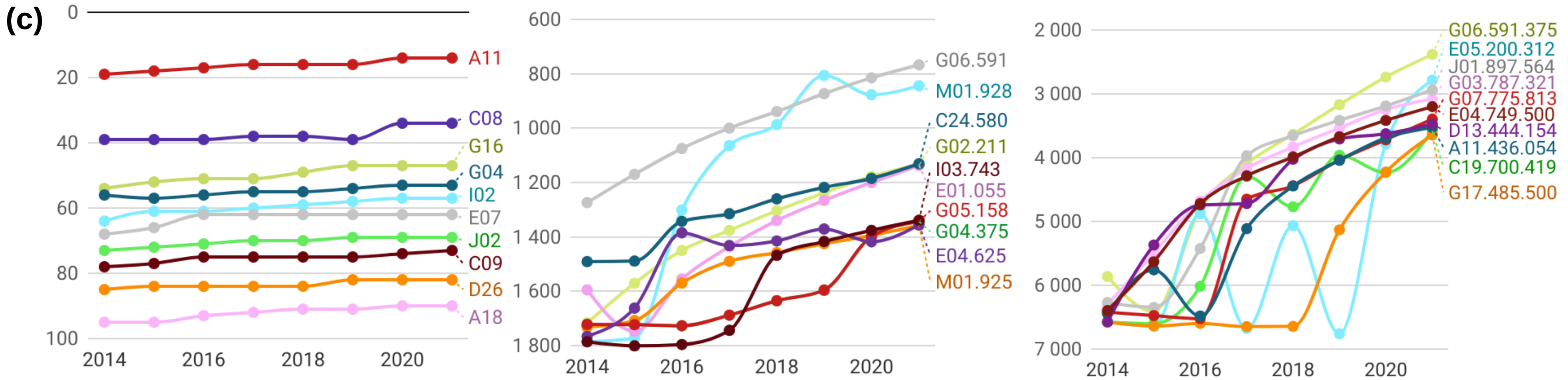}
\includegraphics[width=0.80\linewidth]{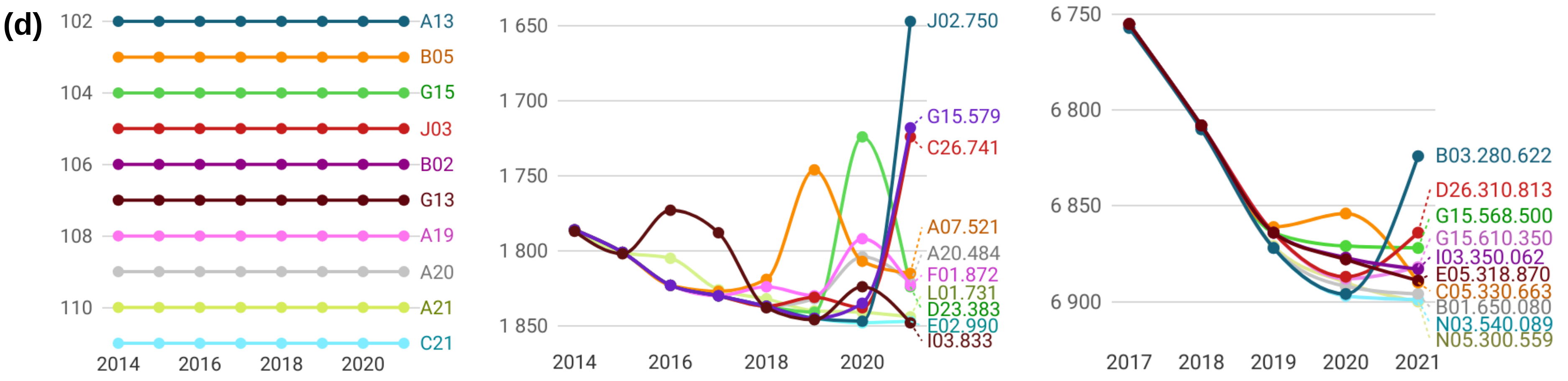}
\caption{Top/bottom concepts over time per level 2 (left), 3 (center), and 4 (right). The left side of each figure shows the rank position of the concept. a) The top 10 concepts with the highest average rank. b) Concepts with the highest rank trend slope. c) Concepts with the lowest rank trend slope. d) The bottom 10 concepts with the lowest average rank. 
} 
\label{figure:top_bottom_concepts_levels}
\end{figure}

Note that some rank shifts reflect relative movements; small score differences can translate into large changes in rank due to the discretization of values.
Figure \ref{figure:top_bottom_concepts_levels}a displays the top 10 concepts according to the concept relevance, at level 2 (left side), level 3 (center), and level 4 (right). 
Half of the top concepts maintain their ranking over time. Concept G02 (Chemical Phenomena) shows variation among several positions, starting in fourth place in 2014 and ending in eighth in 2021. Meanwhile, C12 (Urogenital Diseases), D01 (Inorganic Chemicals), D02 (Organic Chemicals), and C01 (Infections) change by one position over time.

In level 3, four concepts maintain a stable ranking, with B01.050 (Animals) leading. Concepts E05.481 (In Vitro Techniques), A03.620 (Liver), E01.789 (Prognosis), and N06.850 (Public Health) move among multiple positions, with `Prognosis' tending to climb positions over time. In level 4, only the top three concepts, M01.060.116 (Adult), M01.060.406 (Child), and M01.060.057 (Adolescent), remain stable over time. These concepts continue to generate high research interest. The remaining concepts mainly decrease by one position in the ranking.

We computed the rank trend of concepts by adding the differences between consecutive rank positions of a concept and dividing them by the number of differences, i.e., the number of years minus 1. Thus, we identified the concepts with the highest and lowest slopes. The top 10 upward concepts are shown in Figure \ref{figure:top_bottom_concepts_levels}b, which seem to gain positions over time at levels 2, 3, and 4. The behavior of the rank trend of concepts in these levels is similar. The concepts with the highest slope in level 2 are D06 (Hormones, Hormone Substitutes, and Hormone Antagonists), I01 (Social Sciences), D05  (Macromolecular Substances), and C24 (Occupational Diseases). The concept with the highest slope in level 3 is D01.930 (Thorium Compounds), and in level 4 is D03.383.517 (Oxathiins). The downward-trending concepts are shown in Figure \ref{figure:top_bottom_concepts_levels}c, which appear to lose positions over time, in contrast to the upward concepts. The concepts with the lowest rank trend slope are G16 (Biological Phenomena) and I02 (Education), both in level 2. In level 3, the concept M01.928 (Vegetarians) has the lowest trend slope, while in level 4, the concept E05.200.312 (COVID-19 Testing) has the lowest slope.

The bottom concepts are shown in Figure \ref{figure:top_bottom_concepts_levels}d. Concepts in level 2 remain stable, while in levels 3 and 4, concepts change positions constantly. On the other hand, bottom concepts in levels 3 and 4 vary more widely, often by hundreds of ranks, as in these levels, there are more concepts. Most of the changes occur towards the end of the studied period. The concepts J02.750 (`Sustenance,' in level 3) and B03.280.622 (`Pannus,' in level 4) seem to gain positions over time. In Figure \ref{figure:top_bottom_concepts_levels}d, right panel, we are only showing the plot since 2017 for clarity, as previous years follow the trend of 2017-2018. The MeSH headings of the concepts in Figure \ref{figure:top_bottom_concepts_levels} can be found in the MeSH tree~\cite{Meshtree}.

\subsection{Quantitative assessment}
In this section, we describe the quantitative assessment of the multiple relevance aspects through the concept evolution task and retracted papers. We used an ontology evolution dataset for the concept evolution task and identified retracted papers in PubMed. We first describe the datasets, then detail each assessment and present results. We chose the Mann-Whitney test as our data does not follow a normal distribution~\cite{Mann1947}.

\subsubsection{Datasets for quantitative assessment}

\paragraph{Ontology evolution dataset}
We created our dataset for concept evolution as previous studies still need to release their datasets. This dataset contains releases where MeSH descriptors are annotated with the type of change and data-driven temporal features. We calculated these features from the Data-properties dataset (explained in section \ref{subsection:dataset}).

A MeSH descriptor (or concept) is considered to have evolved if it underwent a change in description (change of description), gained new children (extension), changed location (move), or was removed (removal) ~\cite{Cardoso2018}. 
We use the ontology versions of MeSH available in BioPortal, which have two versions per year, e.g., 2014AA and 2014AB releases. When we downloaded the MeSH releases\footnote{Downloaded on April 12, 2022}, releases were available from 2013 to 2021.

We used COntoDiff~\cite{Hartung2013} to get the difference between two consecutive MeSH releases and found no changes in 2013. Thus, we excluded this year. We identified which concepts changed in each release and their type of change (description, extension, move, and removal). 
We used data until 2021AA, as we used 2021AA and 2021AB to identify the concepts that changed in 2021AA. Evolving concepts represent about 1\% of concepts per release, and releases AB have no or very few changes. Over releases, there are $28,900$ concepts on average, the average relevance for evolving concepts is $2.09E-03$, while for concepts that did not evolve is $8.46E-04$.

\paragraph{Retracted papers in PubMed}
We identified the retracted papers in the entire collection of PubMed 2022. Then, we identified the retracted papers in the monthly citation networks. On average, 800 papers were found to be retracted per month. We then retrieved the MeSH annotations of each article, along with the tree nodes associated with each MeSH descriptor. We obtained a value per metric for each article using these tree nodes by summing their metrics’ values.
Since our citation networks are based on a random sample, a paper may appear in multiple months. We therefore averaged each metric per paper per year. We used these values to perform the statistical test.

\subsubsection{Assessment in concept evolution}

We focused on the concept evolution task to investigate whether the approach proposed in this work could contribute to biomedical terminology evolution. We performed a statistical test of significance using the Mann-Whitney test to determine if there is a significant difference between the relevance of descriptors that changed and descriptors that did not change among releases. For the concept relevance (RRF), we found statistical significant differences ($p < 0.05$) whenever evolving descriptors were present.

\begin{figure*}[hbt]
  \centering

  \begin{minipage}[]{0.28\linewidth}
    \centering
    \subcaption{}
    \label{fig:sub-a}
    
    \begin{tabular}{cc}
    \hline
    \textbf{Release} & \textbf{\begin{tabular}[c]{@{}c@{}}Concept \\ relevance\end{tabular}} \\ \hline
    2014AA & \cellcolor{gray!20} $p<0.001$ \\ 
    2014AB & \cellcolor{gray!20} $p<0.001$ \\ 
    2015AA & \cellcolor{gray!20} $p<0.001$ \\ 
    2015AB & \cellcolor{gray!50} $p=0.29$ \\ 
    2016AA & \cellcolor{gray!20} $p<0.001$ \\ 
    2016AB & -- \\ 
    2017AA & \cellcolor{gray!20} $p<0.001$ \\ 
    2017AB & -- \\ 
    2018AA & \cellcolor{gray!20} $p<0.001$ \\ 
    2018AB & \cellcolor{gray!20} $p=0.04$ \\ 
    2019AA & \cellcolor{gray!20} $p<0.001$ \\ 
    2019AB & -- \\ 
    2020AA & \cellcolor{gray!20} $p<0.001$ \\ 
    2020AB & -- \\ 
    2021AA & \cellcolor{gray!20} $p<0.001$ \\ \hline
    \end{tabular}
    
  \end{minipage}
  \begin{minipage}[]{0.50\linewidth}
    \centering
    \subcaption{}
    \label{fig:sub-b}
    \includegraphics[width=0.93\textwidth]{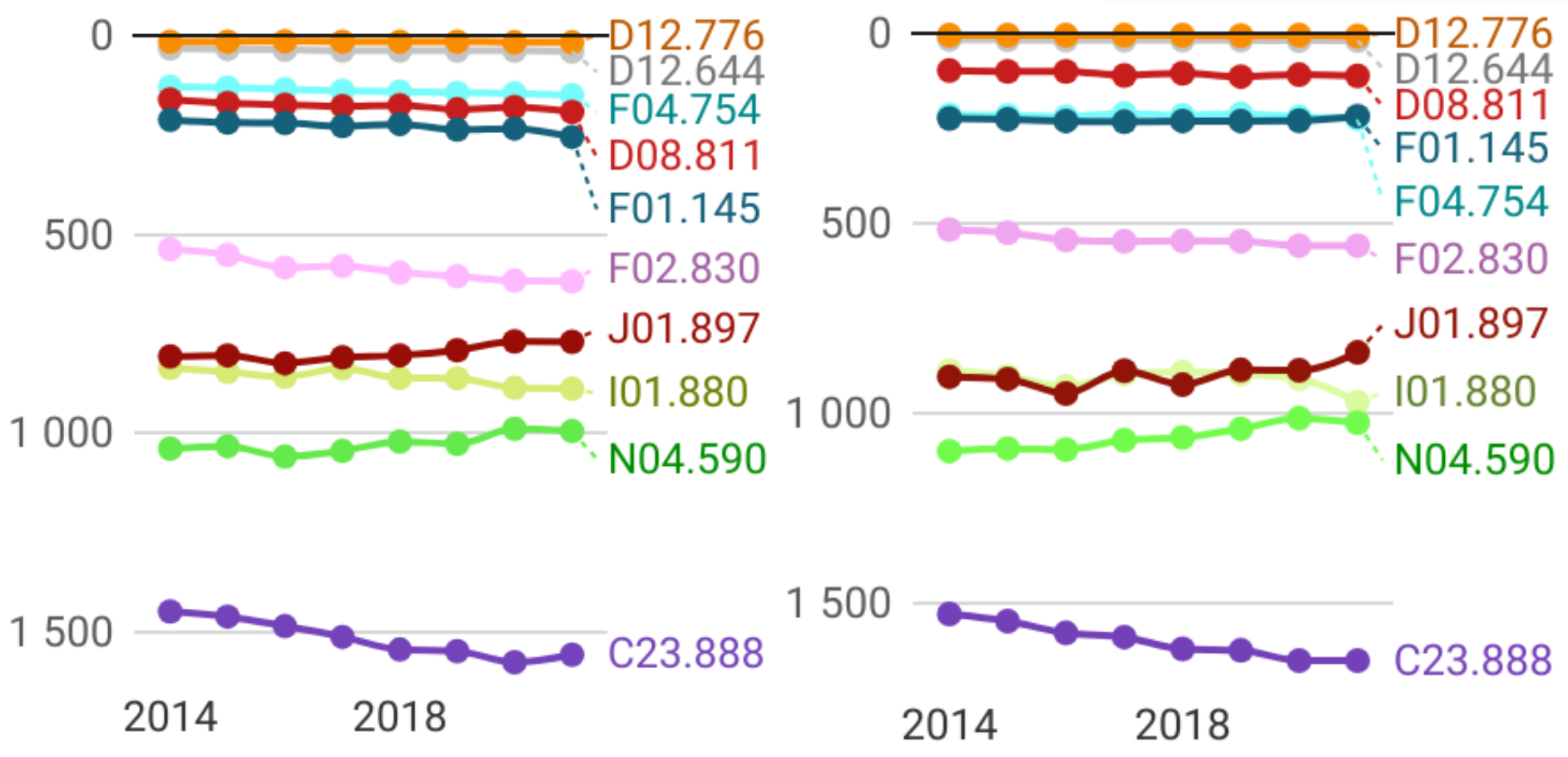}
    \\
    \vspace{.1in}
    \includegraphics[width=0.93\textwidth]{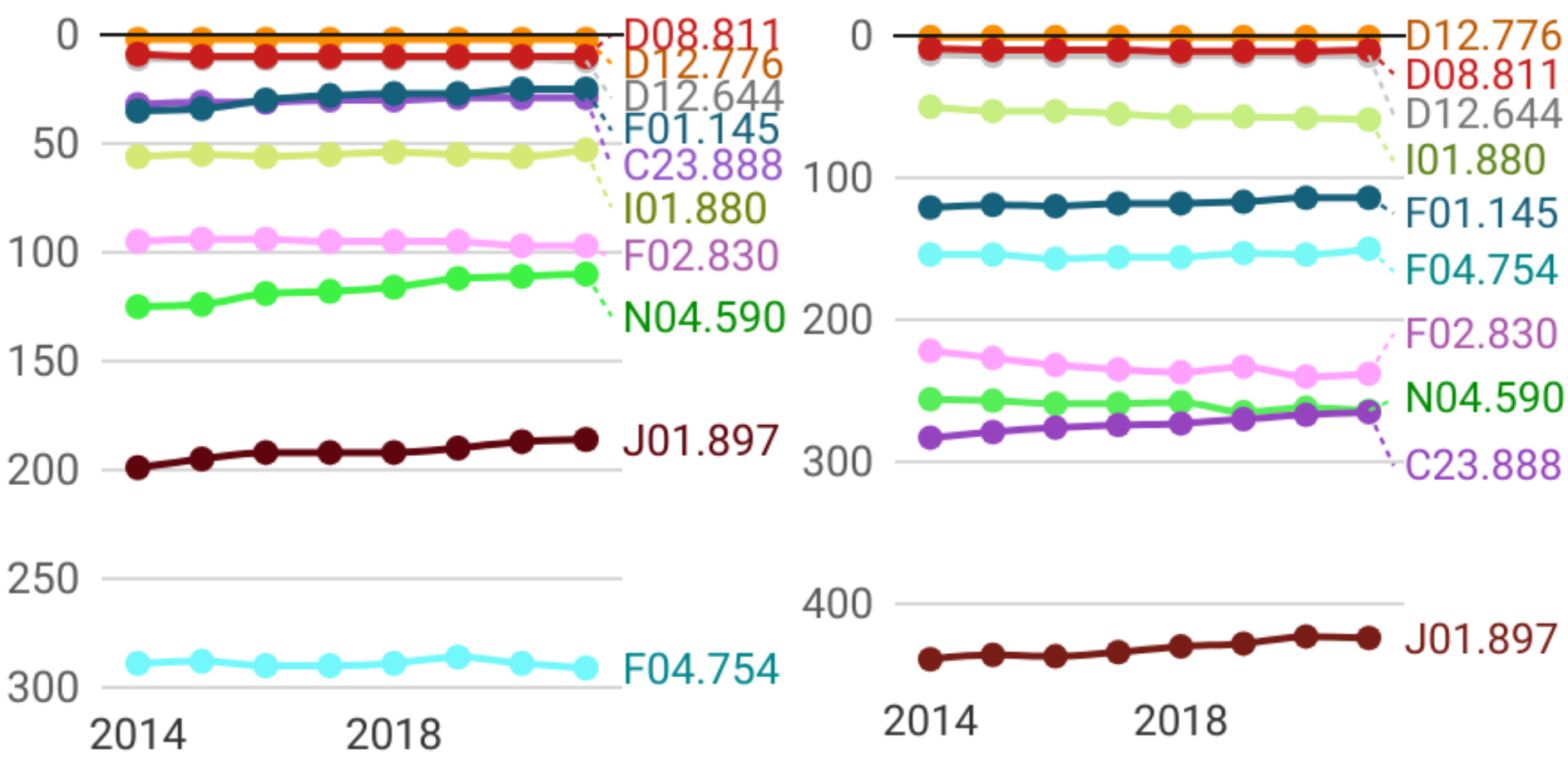}    
  \end{minipage}
  \caption{Evaluation of the multiple aspects in concept evolution. a) Results of the statistical significance test for each release. Light gray-coded cells correspond to $p < 0.05$; otherwise, it is dark gray. b) The top evolving concepts according to each aspect are (starting top left) disruptiveness, influence, informativeness, and usefulness.}
  \label{figure:concept_evolution}
\end{figure*}

Figure \ref{figure:concept_evolution}a shows the test results for each relevance aspect in every release. Releases 2016AB, 2017AB, 2019AB, and 2020AB do not contain evolving descriptors. Consequently, statistical tests could not be applied. Five out of eleven releases exhibit statistically significant differences in usefulness, while in influence and disruptiveness, eight show statistically significant differences. All releases show statistically significant differences in informativeness, but only one release in concept relevance does not demonstrate a statistical difference.

Figure \ref{figure:concept_evolution}b shows the concepts that mostly evolve according to each aspect. Some concepts change their position in disruptiveness and influence. Concepts in informativeness and usefulness remain mostly constant. 
The concepts D12.776 (proteins), D12.644 (peptides), and D08.811 (enzymes) are among the top three aspects. 
The MeSH headings of the remaining concepts can be found in the MeSH tree~\cite{Meshtree}.

\subsubsection{Assessment in retracted papers}

\begin{figure}[hbt]
  \centering
  \begin{minipage}[]{0.50\linewidth}
    \subcaption{}
    \label{fig:sub-a}
    \begin{tabular}{cccccc}
    \hline
    \textbf{Year} & \textbf{\begin{tabular}[c]{@{}c@{}}Disrupti-\\veness\end{tabular}} & \textbf{Influence} & \textbf{\begin{tabular}[c]{@{}c@{}}Informati-\\veness\end{tabular}} & \textbf{Usefulness} & \textbf{\begin{tabular}[c]{@{}c@{}}Concept \\ relevance\end{tabular}} \\ \hline
    2014 & \cellcolor{gray!20}$p=0.02$ & \cellcolor{gray!20}$p<0.001$ & \cellcolor{gray!20}$p<0.001$ & \cellcolor{gray!20}$p<0.001$ & \cellcolor{gray!20}$p<0.001$ \\ 
    2015 & \cellcolor{gray!50}$p=0.68$ & \cellcolor{gray!20}$p<0.001$ & \cellcolor{gray!20}$p<0.001$ & \cellcolor{gray!20}$p<0.001$ & \cellcolor{gray!20}$p<0.001$ \\ 
    2016 & \cellcolor{gray!20}$p=0.03$ & \cellcolor{gray!20}$p=0.002$ & \cellcolor{gray!20}$p<0.001$ & \cellcolor{gray!20}$p<0.001$ & \cellcolor{gray!20}$p<0.001$ \\ 
    2017 & \cellcolor{gray!20}$p<0.001$ & \cellcolor{gray!50}$p=0.56$ & \cellcolor{gray!20}$p<0.001$ & \cellcolor{gray!20}$p<0.001$ & \cellcolor{gray!20}$p<0.001$ \\ 
    2018 & \cellcolor{gray!20}$p<0.001$ & \cellcolor{gray!50}$p=0.69$ & \cellcolor{gray!20}$p<0.001$ & \cellcolor{gray!20}$p<0.001$ & \cellcolor{gray!20}$p<0.001$ \\ 
    2019 & \cellcolor{gray!20}$p<0.001$ & \cellcolor{gray!50}$p=0.38$ & \cellcolor{gray!20}$p<0.001$ & \cellcolor{gray!20}$p<0.001$ & \cellcolor{gray!20}$p<0.001$ \\ 
    2020 & \cellcolor{gray!20}$p<0.001$ & \cellcolor{gray!20}$p<0.001$ & \cellcolor{gray!50}$p=0.15$ & \cellcolor{gray!20}$p<0.001$ & \cellcolor{gray!20}$p<0.001$ \\ 
    2021 & \cellcolor{gray!20}$p<0.001$ & \cellcolor{gray!20}$p<0.001$ & \cellcolor{gray!20}$p=0.05$ & \cellcolor{gray!20}$p<0.001$ & \cellcolor{gray!20}$p<0.001$ \\ \hline
    \end{tabular}

  \end{minipage}
  \hfill
  \begin{minipage}[]{0.28\linewidth}
    \subcaption{}
    \label{fig:sub-b}
    \includegraphics[width=0.9\linewidth]{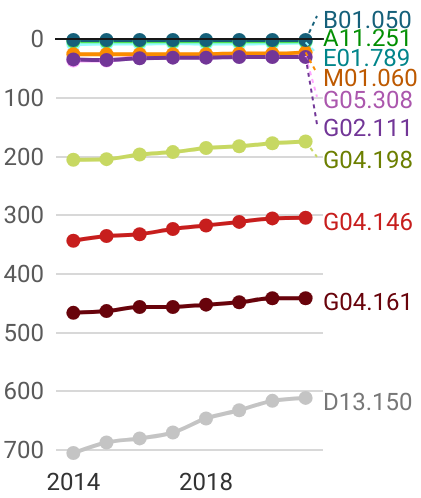}
  \end{minipage}
  \caption{Evaluation of the aspects in retracted papers. a) Results of the statistical significance test. Light gray-coded cells correspond to $p < 0.05$; otherwise, it is dark gray. b) Top-10 concepts in retracted papers according to the concept relevance at level 3.}
  \label{figure:retracted_papers}
\end{figure}

We used the multiple aspects metrics in this study to determine whether there is a significant difference between retracted and non-retracted papers in PubMed. 
The results (Figure \ref{figure:retracted_papers}a) show significant differences between retracted and non-retracted papers in concept relevance, using the proposed aspects as the comparison unit, with a $\textit{p} < 0.05$. Furthermore, when we performed the test by aspect, we also found a statistical difference in usefulness. In influence, there is a significant difference in five out of eight years, while in disruptiveness and informativeness, seven years show differences.

Figure \ref{figure:retracted_papers}b shows the top concepts in retracted papers according to the rank relevance at level 3.
Concepts B01.050 (Animals), A11.251 (Cells, Cultured), and E01.789 (Prognosis) lead the ranking. Half of the concepts come from branch G, and three belong to branch G04 (Cell Physiological Phenomena). The MeSH headings of the remaining concepts can be found in the MeSH tree~\cite{Meshtree}.

\section{Discussion}

Previous research on MeSH has produced useful insights, though with limitations in capturing the multidimensional nature of concept relevance. Structural approaches of hierarchical growth patterns provide insight into organizational principles but do not reflect actual scientific impact or usage patterns~\cite{Balogh2019}. In bibliometric and co-occurrence methods, researchers have attempted to bridge the gap between structure and usage
~\cite{Chien2019,Ilgisonis2022,Kastrin2019}. However, these approaches treat concepts as isolated units without considering citation relationships or hierarchical propagation.
 
Insights from the knowledge graph community offer complementary tools for ranking the importance of classes, i.e., concepts~\cite{Fernandez2021}. These approaches often overlook the rich citation dynamics available in bibliographic resources such as PubMed. Furthermore, studies on scientific disruptiveness indicate that citation patterns can reveal whether a topic displaces prior work or builds on it, insights that raw term counts fail to capture~\cite{Wu2019}. Finally, practical work on annotation and thesaurus maintenance highlights the need to prioritize which concepts to review or update in real-world applications~\cite{Cardoso2018b}. 

A critical gap exists in instance-based analysis, where, despite millions of PubMed articles providing rich instantiation data for MeSH concepts, this resource has remained largely unexploited. Unlike previous work that relies on related or co-occurring descriptors, our approach leverages actual MeSH annotations as concept instantiations, providing a measurement of how concepts are employed in scientific discourse. By integrating information theory with network analysis and hierarchical propagation, our multidimensional framework addresses the limitations of single-perspective approaches, captures temporal evolution, and provides systematic validation through MeSH terminology changes and retracted articles.

Leveraging concepts from information theory (i.e., entropy and category utility) and network analysis (i.e., disruptiveness and graph centrality), we investigated different aspects of the relevance of MeSH concepts. Data instances are propagated through direct mapping for information-theoretical measures and Algorithm \ref{algo:propagation} for network analysis metrics. Direct mapping considers instances when computing these measures.
To summarize these four metrics in an overall concept relevance ranking, we applied the RRF method. In our experiments, we analysed the relevance of MeSH concepts to the PubMed corpus, using the 2022 snapshot annotated with MeSH 2022. Using monthly data from 2014 to 2021, we computed the relevance aspects and showed how they vary over time. We applied the methodology to two use-cases --- concept evolution and retracted articles --- and analyzed the MeSH concept relevance in these scenarios.  

Overall, there is an increasing trend for informativeness and a decreasing trend for usefulness, influence, and disruptiveness. The latter is similar to a recent finding from a study of Park \textit{et al.}, which shows that disruptiveness in life and biomedicine sciences is decreasing ~\cite{Park2023}. At the high level of the MeSH categories, our results show that categories C (\textit{Diseases}) and D (\textit{Chemicals and Drugs}) are the most relevant for the PubMed collection. While this result is expected, given the importance of such concepts in life and biomedical sciences, our method can quantify their relevance according to different aspects, as well as an aggregated relevance rank. In particular, category D ranks first in terms of informativeness and usefulness, while category C leads in influence and disruptiveness. 
The correlation analysis of descriptors revealed a weak correlation among informativeness, usefulness, and disruptiveness, demonstrating that indeed they capture different aspects of concept relevance.

Analysis of the top 10 relevant concepts shows that higher levels are more stable than lower ones due to the propagation of lower-level relevance scores; for instance, level 1 is more stable than level 2. At level 2 (three-digit codes), concepts undergo notable ranking shifts, with Chemical Phenomena (G02) falling by four positions between 2014 and 2021. Levels 3 (six digits) and 4 (nine digits) are even more dynamic: Prognosis (E01.789) rises three positions, while Liver (A03.620) drops three in the same period, and most level 4 concepts lose one position. Within categories C, D, and E, level 2 remains largely stable, but at level 3, Cardiovascular Diseases (C14) and Investigative Techniques (E05) show the greatest changes, possibly reflecting evolving scientific knowledge. Overall, lower-ranked concepts (bottom 10) shift more than higher-ranked ones, except at level 1.

We further evaluated the application of our proposed approach using two analysis use-cases: concept evolution and retracted papers. For the former, we utilized an ontology evolution dataset, containing concept changes from consecutive MeSH versions. Interestingly, the Mann-Whitney test confirmed statistically significant differences between the relevance rank of MeSH concepts that evolve versus those that remain stable. This result suggests that the relevance of a concept could be used to identify candidates for update in KOS maintenance. Additionally, we show that retracted and non-retracted papers can be differentiated using our approach 
even though the monthly proportion of retracted papers was approximately 0.0002\%. In particular, concepts \textit{Animals} (B01.050), \textit{Cells, Cultured} (A11.251), \textit{Prognosis} (E01.789), \textit{Age Groups} (M01.060), \textit{Gene Expression Regulation} (G05.308), and \textit{Biochemical Phenomena} (G02.111) showed consistently high relevance to the retracted corpus. While the former three concepts also appear in the overall top 10 relevance list, the latter three concepts are particular to retracted papers, suggesting potential life and biomedical research areas with issues.

This study has several limitations. First, extending this work to other domains requires a dataset of instances annotated with a KOS, such as PubMed with MeSH. Alternatively, information extraction methods could be used to automatically annotate instances or corpus~\cite{copara2020contextualized,copara2020named,naderi2021ensemble}. Second, all metrics were computed using a sample from the citation network due to the algorithmic complexity of some metrics, which made full computation prohibitive. Finally, metrics are somewhat correlated. 
Within a short period (e.g., one release to the next), our method is not always sensitive enough to identify all changes. 

\section{Conclusions}
In this paper, we present a methodology based on information theory and network analysis to compute the relevance of MeSH concepts for the biomedical corpus indexed by PubMed. By considering multiple relevance aspects of a concept in a data-driven analysis, our approach provides new insights into the evolution of MeSH concept relevance within PubMed. 

We analyzed the dynamics of MeSH over time and showed the potential of our approach in tasks such as tracking concept evolution and identifying retracted papers. We found statistically significant differences in relevance values across these tasks. Future work should extend this methodology to other domains and address the computational complexities associated with some of the metrics.


\begin{credits}
\subsubsection{\ackname} Part of this work has been funded by the CINECA project (H2020 No 825775) and Innosuisse project funding number 46966.1 IP-ICT.

\subsubsection{\discintname}
The authors have no competing interests to declare that are
relevant to the content of this article. 
\end{credits}
%
%
\bibliographystyle{splncs04}
\bibliography{14}

\begin{thebibliography}{10}
\providecommand{\url}[1]{\texttt{#1}}
\providecommand{\urlprefix}{URL }
\providecommand{\doi}[1]{https://doi.org/#1}

\bibitem{Nentidis2021}
Anastasios, N., Anastasia, K., Grigorios, T., Georgios, P.: What is all this new {MeSH} about? International Journal on Digital Libraries  \textbf{22}(4),  319--337 (2021)

\bibitem{Balogh2019}
Balogh, S.G., Zagyva, D., Pollner, P., Palla, G.: Time evolution of the hierarchical networks between pubmed {MeSH} terms. PLoS One  \textbf{14}(8),  e0220648 (2019)

\bibitem{Brin1998}
Brin, S., Page, L.: The anatomy of a large-scale hypertextual web search engine. Computer networks and ISDN systems  \textbf{30}(1-7),  107--117 (1998)

\bibitem{Cardoso2018b}
Cardoso, S.D.: {MAISA - Maintenance of semantic annotations}. Theses, {Universit{\'e} Paris Saclay (COmUE)} (Dec 2018), \url{https://theses.hal.science/tel-02288589}

\bibitem{Cardoso2020}
Cardoso, S.D., Da~Silveira, M., Pruski, C.: Construction and exploitation of an historical knowledge graph to deal with the evolution of ontologies. Knowledge-Based Systems  \textbf{194},  105508 (2020)

\bibitem{Cardoso2018}
Cardoso, S.D., Pruski, C., Da~Silveira, M.: Supporting biomedical ontology evolution by identifying outdated concepts and the required type of change. Journal of Biomedical Informatics  \textbf{87},  1--11 (2018)

\bibitem{Chien2019}
Chien, T.W., Wu, H.M., Wang, H.Y., Chou, W.: The most cited {MeSH} terms and authors who published papers in {Pubmed Central} on the topic of medicine and health using bibliometric analyses. Asian Journal of Medicine and Health  \textbf{14}(4),  1–9 (Apr 2019). \doi{10.9734/ajmah/2019/v14i430105}

\bibitem{copara2020contextualized}
Copara, J., Knafou, J., Naderi, N., Moro, C., Ruch, P., Teodoro, D.: Contextualized french language models for biomedical named entity recognition. In: Actes de la 6e conf{\'e}rence conjointe Journ{\'e}es d'{\'E}tudes sur la Parole (JEP, 33e {\'e}dition), Traitement Automatique des Langues Naturelles (TALN, 27e {\'e}dition), Rencontre des {\'E}tudiants Chercheurs en Informatique pour le Traitement Automatique des Langues (R{\'E}CITAL, 22e {\'e}dition). Atelier D{\'E}fi Fouille de Textes. pp. 36--48 (2020)

\bibitem{copara2020named}
Copara, J., Naderi, N., Knafou, J., Ruch, P., Teodoro, D.: Named entity recognition in chemical patents using ensemble of contextual language models. In: Proceedings of the CLEF 2020 conference. 22-25 September 2020 (2020)

\bibitem{Cormack2009}
Cormack, G.V., Clarke, C.L., Buettcher, S.: Reciprocal rank fusion outperforms condorcet and individual rank learning methods. In: Proceedings of the 32nd international ACM SIGIR conference on Research and development in information retrieval. pp. 758--759 (2009)

\bibitem{Corter1992}
Corter, J.E., Gluck, M.A.: Explaining basic categories: Feature predictability and information. Psychological bulletin  \textbf{111}(2), ~291 (1992)

\bibitem{Fernandez2021}
Fern{\'a}ndez-{\'A}lvarez, D., Frey, J., Labra~Gayo, J.E., Gayo-Avello, D., Hellmann, S.: Approaches to measure class importance in knowledge graphs. Plos one  \textbf{16}(6),  e0252862 (2021)

\bibitem{Figueiredo2019}
Figueiredo, F., Andrade, N.: {Quantifying Disruptive Influence in the AllMusic Guide}. In: {Proceedings of the 20th International Society for Music Information Retrieval Conference}. pp. 832--838. ISMIR (Nov 2019). \doi{10.5281/zenodo.3527940}

\bibitem{Fortunato2018}
Fortunato, S., Bergstrom, C.T., B{\"o}rner, K., Evans, J.A., Helbing, D., Milojevi{\'c}, S., Petersen, A.M., Radicchi, F., Sinatra, R., Uzzi, B., et~al.: Science of science. Science  \textbf{359}(6379),  eaao0185 (2018)

\bibitem{Funk2017}
Funk, R.J., Owen-Smith, J.: A dynamic network measure of technological change. Management Science  \textbf{63}(3),  791--817 (2017). \doi{10.1287/mnsc.2015.2366}

\bibitem{Ghahramani2006}
Ghahramani, Z.: Information Theory. John Wiley \& Sons, Ltd (2006). \doi{10.1002/0470018860.s00643}

\bibitem{gobeill2009question}
Gobeill, J., Pasche, E., Teodoro, D., Veuthey, A.L., Lovis, C., Ruch, P.: Question answering for biology and medicine. In: 2009 9th International Conference on Information Technology and Applications in Biomedicine. pp.~1--5. IEEE (2009)

\bibitem{Gobeill2009TakingBO}
Gobeill, J., Teodoro, D., Pasche, E., Ruch, P.: Taking benefit of query and document expansion using {MeSH} descriptors in medical imageclef 2009. In: Conference and Labs of the Evaluation Forum (2009)

\bibitem{Hamilton2020}
Hamilton, W.L.: Graph representation learning. Morgan \& Claypool Publishers (2020)

\bibitem{Hartung2013}
Hartung, M., Gro{\ss}, A., Rahm, E.: {COnto--Diff}: generation of complex evolution mappings for life science ontologies. Journal of biomedical informatics  \textbf{46}(1),  15--32 (2013)

\bibitem{iCite2022}
iCite, Hutchins, B.I., Santangelo, G., Haque, E.: {iCite Database Snapshot 2022-04}  (5 2022). \doi{10.35092/yhjc.19763386.v1}

\bibitem{Ilgisonis2022}
Ilgisonis, E.V., Pyatnitskiy, M.A., Tarbeeva, S.N., Aldushin, A.A., Ponomarenko, E.A.: How to catch trends using {MeSH} terms analysis? Scientometrics  \textbf{127}(4),  1953--1967 (2022)

\bibitem{Ivanovic2014}
Ivanovi{\'c}, M., Budimac, Z.: An overview of ontologies and data resources in medical domains. Expert Systems with Applications  \textbf{41}(11),  5158--5166 (2014)

\bibitem{Kastrin2019}
Kastrin, A., Hristovski, D.: Disentangling the evolution of {MEDLINE} bibliographic database: A complex network perspective. Journal of biomedical informatics  \textbf{89},  101--113 (2019)

\bibitem{Konopka2015}
Konopka, B.M.: Biomedical ontologies—a review. Biocybernetics and Biomedical Engineering  \textbf{35}(2),  75--86 (2015)

\bibitem{Lipscomb2000}
Lipscomb, C.E.: Medical subject headings. Bulletin of the Medical Library Association  \textbf{88}(3), ~265 (2000)

\bibitem{Lowe1994}
Lowe, H.J., Barnett, G.O.: Understanding and using the {Medical Subject Headings (MeSH)} vocabulary to perform literature searches. JAMA  \textbf{271}(14),  1103--1108 (04 1994). \doi{10.1001/jama.1994.03510380059038}, \url{https://doi.org/10.1001/jama.1994.03510380059038}

\bibitem{Lu2022}
Lu, K., Yang, G., Wang, X.: Topics emerged in the biomedical field and their characteristics. Technological Forecasting and Social Change  \textbf{174},  121218 (2022). \doi{10.1016/j.techfore.2021.121218}

\bibitem{Mann1947}
Mann, H.B., Whitney, D.R.: On a test of whether one of two random variables is stochastically larger than the other. The Annals of Mathematical Statistics  \textbf{18}(1),  50–60 (Mar 1947). \doi{10.1214/aoms/1177730491}

\bibitem{Mizzaro1997}
Mizzaro, S.: Relevance: The whole history. Journal of the American society for information science  \textbf{48}(9),  810--832 (1997)

\bibitem{naderi2021ensemble}
Naderi, N., Knafou, J., Copara, J., Ruch, P., Teodoro, D.: Ensemble of deep masked language models for effective named entity recognition in health and life science corpora. Frontiers in research metrics and analytics  \textbf{6},  689803 (2021)

\bibitem{Nelson2007}
Nelson, S.J., Schulman, J.: A multilingual vocabulary project-managing the maintenance environment. MeSH Section, National Library of Medicine, Bethesda, Maryland  (2007)

\bibitem{Newman2018}
Newman, M.: Networks. Oxford university press (2018)

\bibitem{Meshtree}
NIH: Me{SH} browser, \url{https://meshb.nlm.nih.gov/treeview}, accessed on 14.07.2025

\bibitem{nlm}
NIH: Record types, \url{https://www.nlm.nih.gov/mesh/intro\_record\_types.html}, accessed on 14.07.2025

\bibitem{Park2023}
Park, M., Leahey, E., Funk, R.J.: Papers and patents are becoming less disruptive over time. Nature  \textbf{613}(7942),  138–144 (Jan 2023). \doi{10.1038/s41586-022-05543-x}

\bibitem{Rosch1976}
Rosch, E., Mervis, C.B., Gray, W.D., Johnson, D.M., Boyes-Braem, P.: Basic objects in natural categories. Cognitive Psychology  \textbf{8}(3),  382–439 (Jul 1976). \doi{10.1016/0010-0285(76)90013-x}

\bibitem{Saracevic1975}
Saracevic, T.: Relevance: A review of and a framework for the thinking on the notion in information science. Journal of the American Society for information science  \textbf{26}(6),  321--343 (1975)

\bibitem{Schamber1990}
Schamber, L., Eisenberg, M.B., Nilan, M.S.: A re-examination of relevance: toward a dynamic, situational definition*. Information processing and management  \textbf{26}(6),  755--776 (1990)

\bibitem{Shannon1948}
Shannon, C.E.: A mathematical theory of communication. The Bell system technical journal  \textbf{27}(3),  379--423 (1948)

\bibitem{teodoro2021information}
Teodoro, D., Ferdowsi, S., Borissov, N., Kashani, E., Vicente~Alvarez, D., Copara, J., Gouareb, R., Naderi, N., Amini, P.: Information retrieval in an infodemic: the case of {COVID}-19 publications. Journal of medical Internet research  \textbf{23}(9),  e30161 (2021)

\bibitem{Wang2023}
Wang, X., Kang, H., Fu, L., Yao, L., Ding, J., Wang, J., Gan, X., Zhou, C., Hopcroft, J.E.: Quantifying knowledge from the perspective of information structurization. Plos one  \textbf{18}(1),  e0279314 (2023)

\bibitem{Wu2019}
Wu, L., Wang, D., Evans, J.A.: Large teams develop and small teams disrupt science and technology. Nature  \textbf{566}(7744),  378--382 (Feb 2019)

\end{thebibliography}

\end{document}